%% file: main.tex
\documentclass[journal]{IEEEtran}
\usepackage{amsmath,amssymb,amsfonts,tabularx}
\usepackage[hidelinks]{hyperref}
\usepackage{cite}
\usepackage{url}
\usepackage{paralist}
\usepackage{flushend}
\usepackage{graphicx}
\usepackage{makecell}
\usepackage{xcolor}
\usepackage{array,multirow}
\usepackage{xspace}
\graphicspath{{figures/}}
\usepackage{adjustbox, makecell, colortbl, xcolor}
\usepackage{amssymb}
\usepackage{algorithm}
\usepackage{algpseudocode}
\usepackage{tabularx}
\usepackage{booktabs}
\usepackage{array}
\usepackage{tcolorbox}
\usepackage{placeins}
\usepackage{tikz}
\usetikzlibrary{positioning, arrows.meta, fit, calc}

\usepackage{tikz}
\usetikzlibrary{arrows,arrows.meta,positioning,calc}

\usepackage[table]{xcolor}

\definecolor{lightgray}{gray}{0.9}
\definecolor{verylightblue}{RGB}{230, 245, 255}


\begin{document}
\title{Out of Sync, Out of Sight: Phantom State Attacks against IIoT Intrusion Detection}

\author{Sabrine~Ennaji\textsuperscript{1},
        Elhadj~Benkhelifa\textsuperscript{2},
        and~Nadia~Kabachi\textsuperscript{1}\\[6pt]
\small
\textsuperscript{1}Universit\'e Lyon~1, Universit\'e Lumi\`ere Lyon~2, ERIC, 69100 Villeurbanne, France\\
\textsuperscript{2}School of Computer Science and Mathematics, Liverpool John Moores University, Liverpool, U.K.\\[2pt]
\{sabrine.ennaji, nadia.kabachi\}@univ-lyon1.fr, e.benkhelifa@ljmu.ac.uk
}

\maketitle

\input{abstract}

\IEEEpeerreviewmaketitle

\input{introduction}
\input{background}
\input{framework}
\input{discussion} 
\input{related_work}
\input{conclusion}

\section*{Acknowledgment}
This work was funded by Bpifrance.

\bibliographystyle{IEEEtran}
\bibliography{bibliography}

% \end{IEEEbiography}



\end{document}

%% file: abstract.tex
\begin{abstract}
Machine learning-based intrusion detection systems (IDS) are critical for securing Industrial Internet of Things (IIoT) environments against sophisticated cyber threats. Most adversarial research against these systems perturbs the feature vector, the raw traffic that produces it, or the physical process values reported by field devices, and depends on gradient access, repeated queries to the deployed model, or a learned surrogate or generative model of the victim environment. A smaller line of work reshapes packet timing without querying the detector, but does so by making malicious traffic mimic a learned model of benign timing. Across these approaches, one assumption of industrial monitoring pipelines has received little attention: temporal synchronization. An IDS reconstructs operational state by aggregating distributed telemetry into sliding or tumbling windows, so its view depends not only on what is observed but on when each observation falls relative to a window boundary.

This paper introduces the Phantom State Attack (PSA), which exploits that dependence under a passive, zero-query threat model. Rather than modifying packet contents, perturbing feature values, querying the classifier, or fitting any model of benign traffic, PSA injects bounded timing drift calibrated to the attack flow's own observed inter-arrival variability, applying only the minimal displacement needed to move observations across the nearest aggregation-window boundary. The result is a reconstructed \emph{phantom state} that diverges from the process state that produced it.

We evaluate PSA on two independent industrial datasets, ToN-IoT and the CIC IIoT 2025 (DataSense) benchmark, against Random Forest, MLP, and XGBoost classifiers, measuring detection degradation, synchronization distortion, stealth against four synchronization-aware detectors, and attacker cost against a query-based baseline. PSA degrades detection on flows that carry enough packets for window-boundary redistribution to take effect, and leaves other categories almost unchanged, so its effect is conditional on target-flow traffic characteristics rather than uniform. A query-based baseline reaches higher raw attack success but needs many model queries per window, whereas PSA needs none, a categorical rather than incremental difference in required attacker capability. The results identify temporal aggregation as an attack surface in IIoT intrusion detection that is reachable under assumptions weaker than those of prior evasion techniques.
\end{abstract}

%% file: introduction.tex
\section{Introduction}

Modern critical infrastructures, ranging from automated manufacturing systems to intelligent energy grids, increasingly rely on the Industrial Internet of Things (IIoT) to support real-time monitoring, process automation, and operational decision-making \cite{latif2025hardware, zhu2025area}. As these environments become more interconnected, they also become more exposed to sophisticated cyber threats capable of disrupting critical services and industrial processes \cite{lakhal2025next}. To address this challenge, machine learning (ML)-based Intrusion Detection Systems (IDS) have emerged as a key security mechanism, owing to their ability to identify complex attack patterns and previously unseen threats from network traffic and telemetry data \cite{islam2025ai}.

Despite their effectiveness, the growing adoption of ML-based IDS has introduced new security concerns \cite{ennaji2025adversarial}. Adversarial machine learning has shown that carefully crafted inputs can manipulate IDS decisions and reduce detection effectiveness \cite{ennaji2026vulnerability}. Most of this work perturbs packet contents, protocol fields, or extracted feature vectors while preserving attack functionality \cite{yao2025modeling, verma2025secure, rezaei2025federated}, and in doing so depends on at least one of three capabilities: white-box access to model gradients, repeated black-box interaction with the deployed classifier to refine an evasive input, or a learned surrogate or generative model of the victim environment. Real deployments increasingly undermine the querying capability through rate-limiting, logging, and query-pattern monitoring. A narrower line of work reshapes packet timing alone, without querying the detector, by training a model of benign inter-packet delays and re-timing attack packets to resemble it \cite{sharon2022tantra}. That approach removes the query requirement, but still needs a capture of benign traffic from the target network and a trained sequence model, and it works by mimicry rather than by exploiting the monitoring pipeline itself.

Industrial monitoring pipelines, however, depend on more than the values of individual features. In practical IIoT deployments, network packets arrive as continuous and highly correlated streams of process events \cite{rathi2025realizing}. To process this information, IDS pipelines commonly aggregate raw packet arrivals into sliding or tumbling windows and compute statistical features over each window. The operational view reconstructed by the IDS therefore depends not only on what information is observed, but also on when each observation falls relative to a window boundary. This introduces an implicit assumption: that the temporal placement of telemetry observations faithfully reflects the evolution of the underlying industrial system.

The security implications of this assumption remain largely unexplored. If an attacker can subtly shift communication timing while preserving protocol semantics and traffic plausibility, the resulting telemetry stream stays syntactically valid yet becomes temporally inconsistent. Such inconsistency changes which observations fall in which window, which changes the statistical features extracted from them, which in turn can change the state the IDS reconstructs \cite{ennaji2025adversarial}. This manipulation requires no interaction with the IDS: an attacker that controls only the scheduling of its own packets, with no read access to the model, its features, or its decisions, can still influence the operational view the IDS reconstructs, so that malicious activity may remain active while appearing less suspicious.

To investigate this surface, we introduce the \emph{Phantom State Attack} (PSA), a synchronization-based adversarial attack against ML-based IIoT intrusion detection under a passive, zero-query threat model. Rather than perturbing packet contents, manipulating feature values, querying the classifier, or fitting any model of benign traffic, PSA injects bounded timing drift, calibrated to the attack flow's own observed inter-arrival variability, to move observations across the nearest aggregation-window boundary. By disrupting temporal coherence at the point of aggregation, the attack causes the IDS to reconstruct a distorted operational representation, the \emph{phantom state}, that no longer reflects the true state of the monitored process.

Unlike many adversarial studies that focus on a single target architecture or a single evaluation setting, this work asks whether a synchronization-based vulnerability generalizes across learning paradigms, attack categories, and independent industrial testbeds. PSA is evaluated under a black-box, zero-query setting using Random Forest (RF), XGBoost, and Multi-Layer Perceptron (MLP) classifiers across attack categories spanning the ToN-IoT and CIC IIoT 2025 (DataSense) datasets. The evaluation analyzes detection degradation, attack success rate, synchronization distortion, and stealth under multiple drift budgets, and adds a direct comparison against a query-based baseline that isolates attacker cost, measured in required model interactions. The results show that timing manipulation degrades detection on flows carrying enough packets for window-boundary redistribution to take effect, that the effect is conditional on target-flow characteristics rather than uniform, and that where it occurs it is achieved at a categorically lower attacker cost than query-based feature-space attacks.

The main contributions of this paper are as follows:

\begin{itemize}

\item We identify temporal aggregation, specifically the placement of observations relative to sliding or tumbling window boundaries, as an attack surface in ML-based IIoT intrusion detection that is reachable under a passive, zero-query threat model weaker than prior query-based evasion.

\item We propose the Phantom State Attack (PSA), which exploits this surface through flow-calibrated, minimally-sufficient timing drift, using no surrogate model and no model of benign traffic.

\item We evaluate PSA across attack categories on the ToN-IoT and CIC IIoT 2025 (DataSense) datasets using RF, MLP, and XGBoost, analyzing detection degradation, attack success, and synchronization distortion under multiple drift budgets, and we report the conditions under which the attack does and does not transfer.

\item We assess PSA's stealth against four synchronization-aware detectors and compare its attacker cost and traffic consistency against a query-based baseline, quantifying the trade-offs inherent to each threat model.

\end{itemize}

The remainder of this paper is organized as follows. Section~\ref{sec:background} presents the background on temporal synchronization and industrial monitoring pipelines. Section~\ref{sec:framework} introduces the proposed Phantom State Attack (PSA), including the threat model and attack workflow. Section~\ref{sec:results} presents the experimental evaluation and discussion. Section~\ref{sec:related} positions PSA with respect to prior work on adversarial attacks against ML-based IDS, timing-based traffic manipulation, and synchronization attacks on industrial infrastructure. Finally, Section~\ref{sec:conclusion} concludes the paper and outlines future research directions.

%% file: background.tex
\section{Background and Technical Foundations}
\label{sec:background}
\begin{figure*}[!t]
\centering
\includegraphics[width=0.85\textwidth]{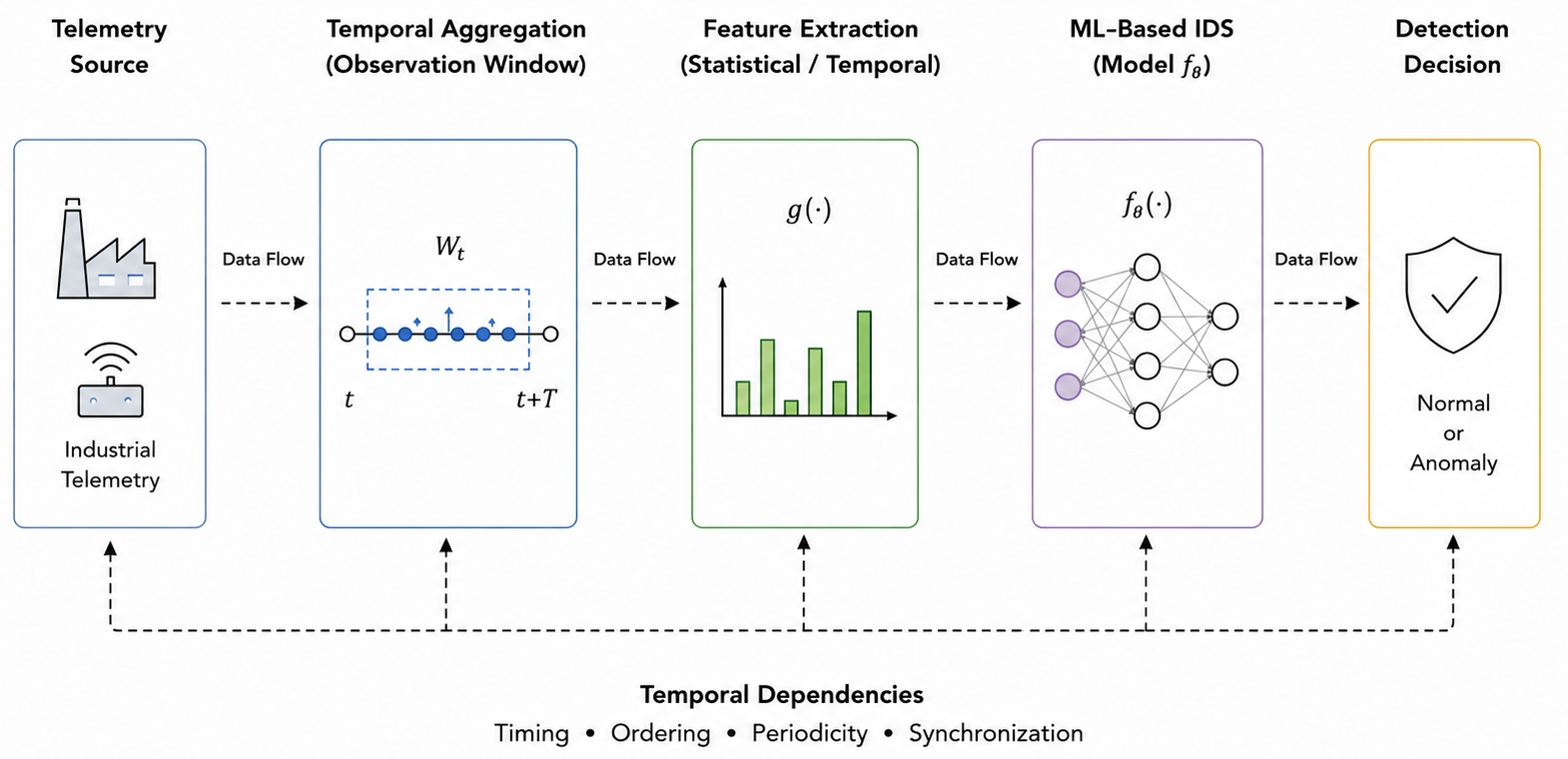}
\caption{General architecture of an ML-based intrusion detection system in IIoT environments. Industrial telemetry is aggregated into temporal observation windows (\(W_t\)), transformed into feature representations through a feature extraction process (\(g(\cdot)\)), and analyzed by a machine learning model (\(f_{\theta}\)) to generate detection decisions.}
\label{fig:iiot_ids_pipeline}
\end{figure*}

This section presents the technical foundations underlying the proposed Phantom State Attack (PSA). We first introduce the general architecture of machine learning-based intrusion detection systems (IDS) in Industrial Internet of Things (IIoT) environments, then describe how industrial monitoring pipelines reconstruct operational behavior through temporal feature extraction, and finally discuss why temporal desynchronization constitutes a previously underexplored adversarial attack surface.

\subsection{ML-Based Intrusion Detection in IIoT Environments}

Industrial Internet of Things (IIoT) environments consist of interconnected sensors, programmable logic controllers (PLCs), industrial gateways, edge devices, and cloud services that continuously exchange operational telemetry and communication data \cite{malik2021industrial}. The increasing scale and complexity of these environments have made traditional signature-based intrusion detection insufficient to identify sophisticated and previously unseen cyber threats \cite{zolanvari2019machine}.

To address this challenge, ML-based IDS have become increasingly integrated into industrial cybersecurity architectures \cite{jayalaxmi2023pignus}. Unlike signature-based approaches, ML-based IDS learn behavioral patterns from historical observations and use these patterns to distinguish benign from malicious activities \cite{sahani2023machine}. As illustrated in Figure~\ref{fig:iiot_ids_pipeline}, the final detection decision is produced by a trained machine learning model operating on extracted traffic features.

Formally, let

\[
x_t \in \mathbb{R}^{d}
\]

denote the feature representation extracted at time (t), where (d) represents the number of extracted features. The IDS prediction is given by

\[
y_t=f_{\theta}(x_t),
\]

where \(f_{\theta}\) denotes the trained detection model and \(y_t\) represents the predicted class label.

Hence, the effectiveness of the IDS depends on the quality and representativeness of the extracted feature representation supplied to the classifier.

\subsection{Temporal Feature Reconstruction in Industrial Monitoring Systems}

Industrial communication processes are inherently time-dependent \cite{barbieri2025close}. Sensors periodically transmit telemetry measurements, controllers exchange synchronized operational messages, and monitoring infrastructures continuously collect observations to reconstruct the state of the monitored process.

Rather than independently analyzing individual packets, most ML-based IDS pipelines aggregate observations into temporal windows prior to feature extraction. This approach enables the system to capture communication dynamics such as packet rates, burst behavior, timing characteristics, and protocol activity patterns that are difficult to observe at the packet level.

Let

\[
W_t={p_i \mid t \leq t_i < t+T}
\]

denote a temporal aggregation window of duration \(T\), where \(p_i\) represents the \(i\)-th packet or telemetry observation occurring at time \(t_i\).

The corresponding feature representation is obtained as

\[
x_t=g(W_t),
\]

where \(g(\cdot)\) presents the feature extraction process that transforms raw observations into statistical and temporal features used by the IDS.

Let

\[
S(t)
\]

describes the true operational process state at time (t), and 

\[
\hat{S}(t)
\]

represent the state reconstructed by the IDS from observed telemetry.

Under synchronized operation,

\[
\hat{S}(t)\approx S(t),
\]

indicating that the reconstructed state accurately reflects the evolution of the underlying process. Therefore, temporal aggregation enables the IDS to capture communication dynamics and reconstruct operational behavior from continuously evolving telemetry streams.

\begin{figure*}[!t]
\centering
\includegraphics[width=0.85\textwidth]{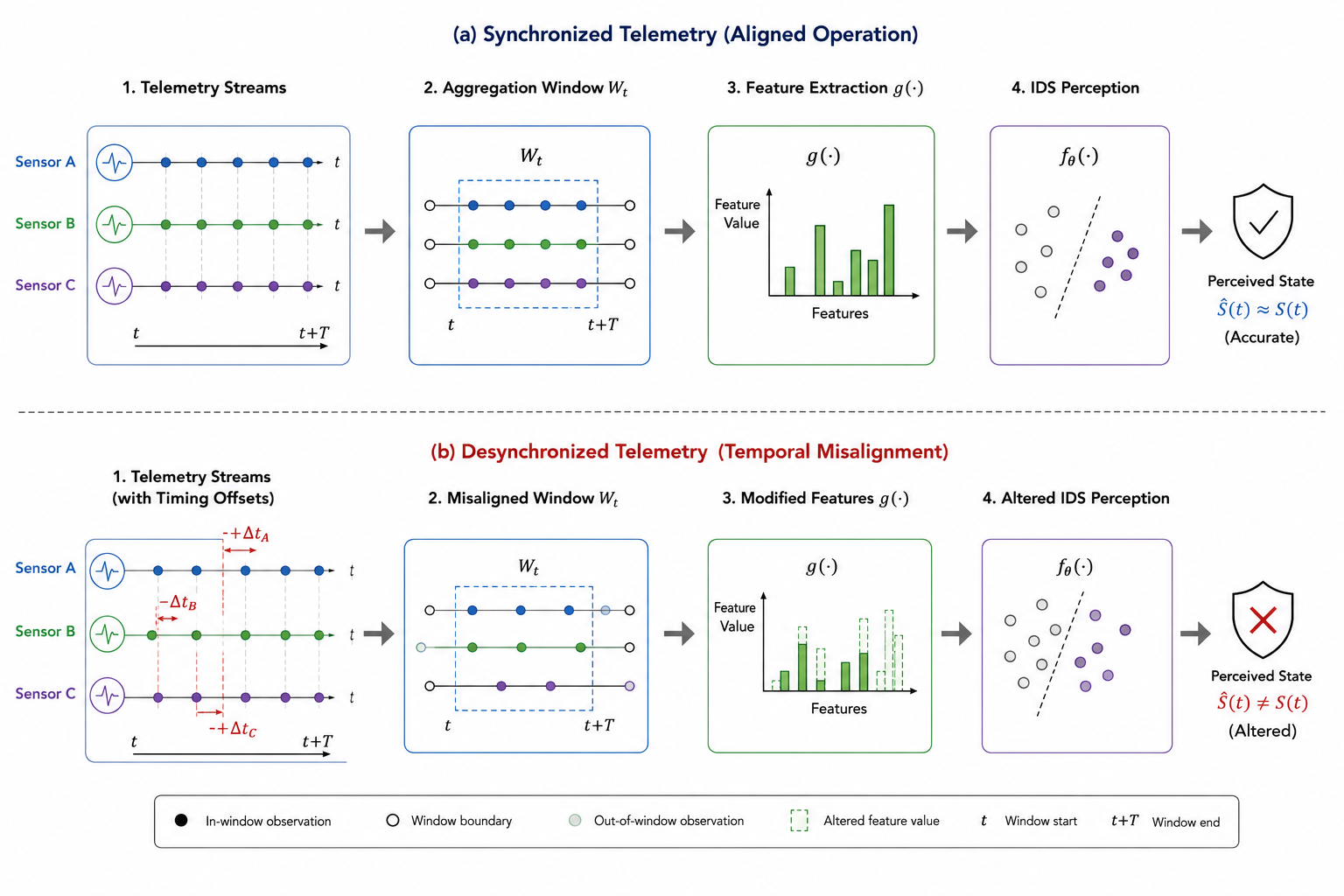}
\caption{Temporal synchronization dependency in industrial monitoring systems. Timing offsets across telemetry streams may alter aggregation-window composition, modify extracted features, and affect the state reconstructed by the IDS.}
\label{fig:temporal_pipeline}
\end{figure*}

\subsection{Temporal Desynchronization as an Adversarial Attack Surface}

Most adversarial attacks against IDS focus on directly manipulating packet attributes, payload values, or extracted features to influence classifier decisions \cite{ennaji2025adversarial}. Although these approaches reveal important weaknesses in ML-based IDS, they generally assume that the primary attack surface lies within the feature space itself \cite{ennaji2026vulnerability}.

However, the temporal aggregation process described in the previous subsection introduces an additional dependency that has received comparatively little attention: temporal consistency. Because feature representations are derived from observations grouped within aggregation windows, changes in observation timing may influence the information supplied to the classifier.

Consequently, an adversary does not necessarily need to alter packet contents to affect detection outcomes. By introducing carefully controlled timing variations, observations may be redistributed across adjacent aggregation windows while remaining protocol-valid and operationally plausible. Although the communication content remains unchanged, the resulting feature representations can differ significantly from those generated under synchronized conditions.

This creates a new adversarial attack surface based on temporal desynchronization rather than direct feature manipulation. In such scenarios, the IDS reconstructs an operational representation that no longer accurately reflects the true evolution of the monitored process. The resulting divergence between the actual process state and the IDS-perceived state forms the foundation of the Phantom State Attack (PSA) presented in the next section.

While the general vulnerability described above extends to desynchronization across multiple independently-timed telemetry sources (as illustrated in Figure~\ref{fig:temporal_pipeline}), the PSA implementation and evaluation presented in this work focus on the single-flow case, in which an attacker perturbs the timing of packets within a single communication flow relative to a fixed aggregation-window boundary; multi-source cross-sensor desynchronization is discussed as a direction for future work.

%% file: framework.tex
\section{Phantom State Attack Methodology}
\label{sec:framework}

This section presents the proposed Phantom State Attack (PSA) and its evaluation methodology. First, we define the threat model and attack assumptions under a passive, zero-query black-box setting. Next, we describe the PSA framework and its flow-calibrated temporal desynchronization strategy for exploiting synchronization dependencies within industrial monitoring pipelines. Finally, we present the experimental setup used to assess the effectiveness, stealth, and operational feasibility of the proposed attack.

Section~\ref{sec:related} and Table~\ref{tab:related} place PSA among prior adversarial attacks on two axes: the object the attacker manipulates, and the interaction the attacker requires with the deployed detector. Read against that landscape, PSA occupies a configuration that prior work does not: it perturbs timing only, it requires no interaction with the detector, and it fits no surrogate model and no model of benign traffic. The attacker constructs the phantom state from the attack flow's own observed timing. Query-based evasion, by contrast, requires an oracle: the attacker submits candidate inputs to the deployed system and observes its output, often many times per window, a cost Section~\ref{sec:results} quantifies empirically. Timing-reshaping attacks that avoid the oracle instead train a model of benign timing and make the attack mimic it, which PSA also avoids.

\subsection{Threat Model and Attack Assumptions}

The considered threat model specifies the target environment and the attacker's capabilities within an IIoT monitoring scenario.

\subsubsection{Target Environment}

We consider an adversary targeting a machine learning-based intrusion detection system deployed within an IIoT monitoring environment. The IDS continuously monitors industrial communication traffic and analyzes telemetry observations to classify operational behavior as benign or malicious.

To evaluate whether synchronization-oriented vulnerabilities generalize across heterogeneous learning paradigms, the monitored IDS architecture includes three different machine learning models:

\begin{itemize}
\item \textbf{Random Forest (RF):} represents tree-based ensemble intrusion detection approaches widely used in industrial IDS environments due to their robustness and low computational complexity \cite{resende2018survey}.
\item \textbf{Multi-Layer Perceptron (MLP):} represents neural-based intrusion detection architectures capable of learning non-linear traffic behavior and temporal correlations \cite{zhao2025research}.
\item \textbf{XGBoost:} represents gradient-boosting intrusion detection models known for high detection performance and sensitivity to complex feature interactions \cite{bhati2021improved}.
\end{itemize}

The use of these heterogeneous IDS architectures allows evaluating whether PSA-induced synchronization distortion affects multiple learning paradigms rather than targeting a single classifier family only.

\subsubsection{Attacker Capabilities and Assumptions}

The attacker operates under a black-box, \emph{zero-query} setting. The adversary has no access to the IDS architecture, model parameters, training data, gradients, or internal detection thresholds, and submits no queries to the deployed model at any stage. This separates PSA from query-based black-box evasion, which lacks white-box access but still requires iterative interaction with the target to construct a perturbation.

The attacker is assumed to control the scheduling of its own attack flow and to passively observe its own communication timing, including inter-arrival patterns and burst structure. Controlling the transmission schedule of traffic one originates is weaker than the capabilities most evasion attacks assume: it requires no read access to the model, no feature-space write access, and no benign-traffic corpus from the protected network. It is available to any adversary that already originates the malicious flow, for example from a compromised device on the local segment. The primary attack therefore schedules the flow's own packets, which allows a packet to be advanced or delayed relative to a window boundary. We separately evaluate a strictly weaker delay-only variant, in which packets may only be delayed, never advanced, modelling an in-path element that can hold a packet but cannot send it earlier (Section~\ref{sec:results}).

One quantity in the pipeline is not strictly an attacker secret but is also not published by the IDS: the window duration $T$ and origin $t_0$ used for aggregation. PSA computes its minimal-sufficient shift against an assumed $(T, t_0)$. Because common feature-extraction tools default to a small set of window sizes, an attacker can approximate these parameters from observed traffic periodicity, and we do not assume they are known exactly. Section~\ref{sec:results} reports a sensitivity analysis in which the attacker assumes the wrong $T$ and a shifted $t_0$ while the IDS aggregates with the true values, to establish how much the attack depends on this approximation.

These manipulations alter the temporal placement of telemetry observations while preserving protocol-valid communication behavior; packet contents, order within the flow, and payloads are unchanged.

\subsection{Phantom State Attack Framework}

The proposed Phantom State Attack (PSA) exploits synchronization dependencies within industrial monitoring pipelines through controlled, flow-calibrated temporal desynchronization. Rather than modifying packet contents, directly perturbing feature values, or querying the deployed classifier, PSA manipulates the timing of communication events to influence how industrial telemetry is aggregated and interpreted by the intrusion detection system.

Based on the synchronization dependency described in Section~\ref{sec:background}, PSA introduces bounded temporal drift into selected communication flows. Unlike a naive strategy that injects uniformly random drift over a fixed range, PSA calibrates the injected drift to each flow's own observed timing variability, and applies only the minimal shift required to move a packet across the nearest aggregation-window boundary, augmented with bounded random jitter drawn from the flow's remaining plausible timing budget. This design serves two purposes. First, the injected drift stays within the flow's own baseline variability, so it does not introduce an artificial, uniformly-distributed timing signature. Second, it maximizes packet redistribution for a given budget, since by default only the boundary-crossing displacement, not the full budget, is spent on each packet. Section~\ref{sec:results} compares PSA directly against the naive uniform-drift strategy to isolate the contribution of boundary-targeted calibration from that of timing perturbation in general.

The PSA workflow consists of four sequential stages:

\begin{enumerate}
\item \textbf{Communication Timing Observation:} The attacker passively observes communication timing characteristics, including telemetry periodicity, packet inter-arrival patterns, and burst behavior.
\item \textbf{Sensitive Flow Selection:} Based on the observed timing behavior, synchronization-sensitive communication flows are identified as candidate targets, namely flows exhibiting sufficient packet volume and measurable inter-arrival regularity.
\item \textbf{Flow-Calibrated Drift Injection:} For each packet in a selected flow, the minimal shift required to reach the nearest window boundary is computed; if this shift falls within the flow's own natural timing variance, bounded jitter is added and the drift is applied.
\item \textbf{Phantom State Induction:} The temporally manipulated traffic is processed by the IDS, resulting in modified feature reconstruction and a distorted operational perception.
\end{enumerate}

To preserve operational realism, the injected temporal drift for a packet $p_i$ in flow $f$ must satisfy

\[
\Delta t_{\min}(p_i) \leq \Delta t(p_i) \leq \beta_f,
\qquad
\beta_f = \min\!\big(\sigma_{\text{IAT}}(f),\, \Delta_{\max}\big),
\]

where $\Delta t_{\min}(p_i)$ is the minimal displacement that moves $p_i$ across its nearest window boundary, $\sigma_{\text{IAT}}(f)$ is the flow's own inter-arrival-time standard deviation, which bounds plausible timing variation for that flow, and $\Delta_{\max}$ is a global drift budget that the evaluation varies. The per-flow budget $\beta_f$ is thus the tighter of the two: a flow is never drifted beyond its own natural variability, and never beyond the global budget. A packet whose boundary-crossing shift exceeds $\beta_f$ is left unmodified, so flows with little natural timing variation are perturbed little or not at all.

When a packet is shifted, it is redistributed into an adjacent window without altering its content or its order within the flow. The window features the IDS computes then differ from those it would compute under synchronized conditions, and the monitoring system reconstructs the distorted operational perception we term a \emph{phantom state}.

\begin{figure*}[!t]
\centering
\includegraphics[width=0.95\textwidth]{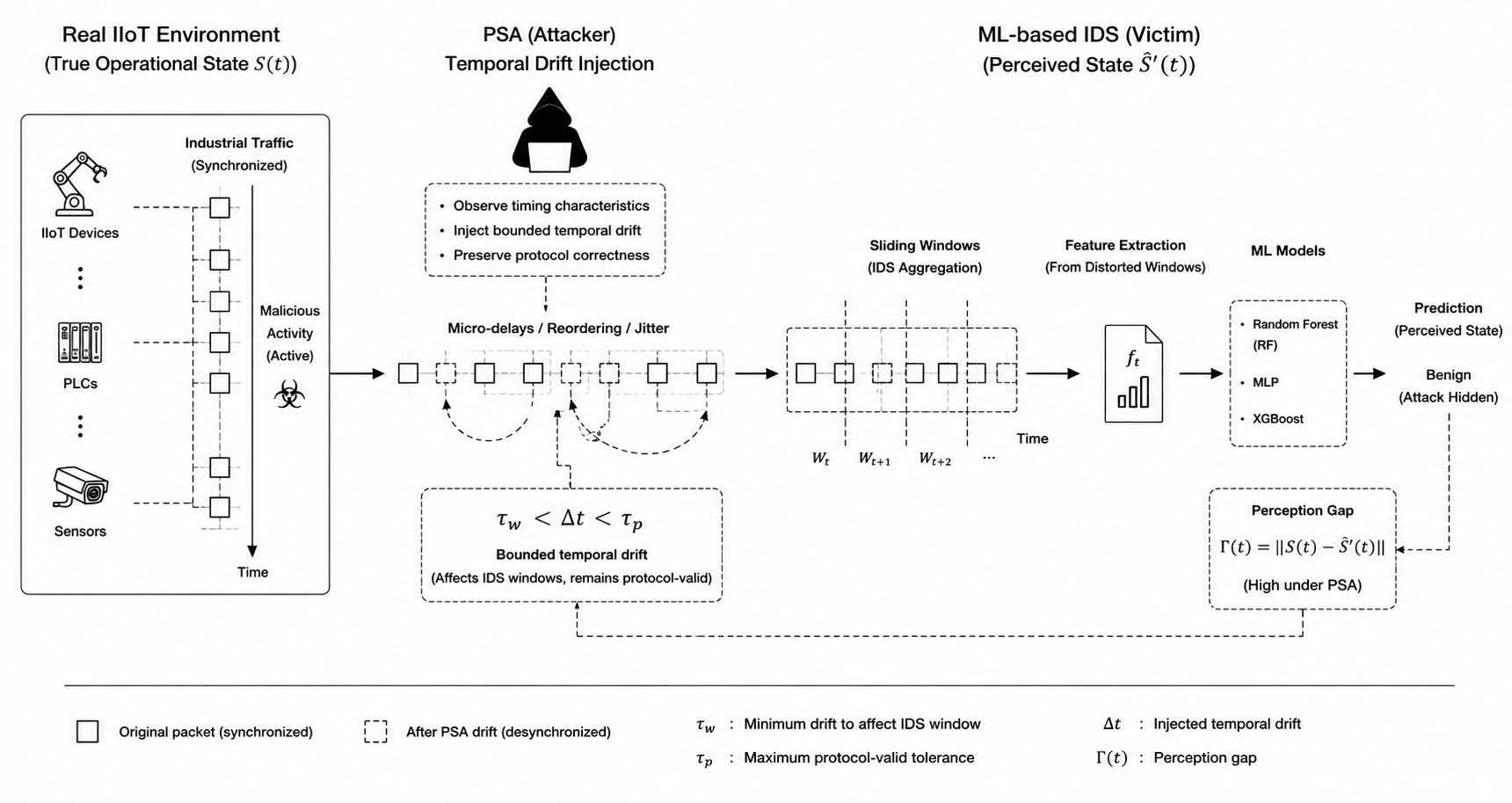}
\caption{Phantom State Attack workflow. The attacker passively observes the timing of its own flow, selects synchronization-sensitive flows, injects flow-calibrated minimal-sufficient drift to move packets across the nearest aggregation-window boundary, and the IDS then reconstructs a distorted phantom state from the re-aggregated windows. Packet contents and order within the flow are unchanged.}
\label{fig:psaframework}
\end{figure*}

Figure~\ref{fig:psaframework} illustrates the overall PSA workflow, while Algorithm~\ref{alg:psa} summarizes the corresponding attack procedure.

\begin{algorithm}[!t]
\caption{Phantom State Attack (PSA): Jittered Minimal-Sufficient Drift Injection}
\label{alg:psa}
\begin{algorithmic}[1]
\Require Telemetry stream $P$, assumed window duration $T$, assumed origin $t_0$, global budget $\Delta_{\max}$
\Ensure Temporally manipulated telemetry stream $P'$

\State Observe telemetry periodicity, burst behavior, and packet inter-arrival patterns from $P$
\State Identify synchronization-sensitive flows $F_s \subseteq F(P)$: flows with sufficient packet volume and measurable inter-arrival regularity
\State Initialize $P' \leftarrow P$

\For{each flow $f \in F_s$}
    \State Compute flow-specific budget $\beta_f \leftarrow \min(\sigma_{\text{IAT}}(f),\, \Delta_{\max})$
    \For{each packet $p_i \in f$}
        \State Compute distance to nearest window boundary: $d_i \leftarrow \min(t_i \bmod T,\; T - (t_i \bmod T))$
        \State Compute minimal-sufficient shift: $\Delta t_{\min} \leftarrow d_i + \epsilon$
        \If{$\Delta t_{\min} \leq \beta_f$}
            \State Sample jitter $\delta \sim \mathcal{U}(0, \beta_f - \Delta t_{\min})$
            \State $\Delta t \leftarrow \Delta t_{\min} + \delta$
            \State Apply signed drift $t_i' \leftarrow t_i \pm \Delta t$ toward the nearest boundary (delay-only variant: $+$ only)
        \Else
            \State Leave $p_i$ unmodified
        \EndIf
    \EndFor
\EndFor

\State Re-derive window assignment for all packets under updated timestamps
\State \Return manipulated telemetry stream $P'$
\end{algorithmic}
\end{algorithm}

\subsection{Operational Feasibility and Experimental Setup}

The objective of the experimental evaluation is to assess whether flow-calibrated temporal desynchronization can influence intrusion detection performance while preserving realistic industrial communication behavior, and to quantify the attacker-capability trade-offs relative to conventional evasion techniques. Beyond measuring classification degradation, the evaluation investigates how synchronization-aware feature reconstruction is affected when telemetry observations become progressively displaced from their original temporal positions, and how PSA's operational realism and required attacker interaction compare against representative baseline attacks.

\subsubsection{Experimental Datasets}

To evaluate the robustness of PSA across heterogeneous industrial communication environments, experiments are conducted using two representative industrial intrusion detection datasets, spanning fourteen attack categories in total. Table~\ref{tab:datasets} summarizes the datasets employed throughout the evaluation.

\begin{table*}[!t]
\centering
\caption{Industrial datasets used for PSA evaluation.}
\label{tab:datasets}
\begin{tabular}{p{3cm}p{6cm}p{6cm}}
\toprule
\textbf{Dataset} & \textbf{Description} & \textbf{Industrial Characteristics} \\
\midrule

ToN-IoT &
Heterogeneous IoT/IIoT dataset providing both raw packet captures and Zeek/Bro-derived netflow logs with explicit per-record timestamps (\texttt{ts}), collected from normal and malicious telemetry across networked IoT and industrial environments &
Fine-grained, millisecond-level timestamp fields enabling inter-arrival-time computation and controlled temporal drift injection at sub-window resolution; seven evaluated attack categories (backdoor, DoS, DDoS, scanning, MITM, ransomware, password) representative of industrial communication environments \\
\midrule

CIC IIoT Dataset 2025 (DataSense)~\cite{firouzi2025datasense} &
Recent industrial IDS benchmark built on a dedicated 40-device CIC/UNB testbed, providing synchronized sensor and network traffic streams across IoT/IIoT, edge, and cloud layers using real industrial protocols (Modbus TCP, OPC UA, MQTT) &
Shorter, single-session captures with per-packet timestamps, providing an independent testbed whose traffic characteristics differ from ToN-IoT; seven evaluated attack categories (DoS, DDoS, malware, MITM, web-exploitation, brute-force, reconnaissance) representative of contemporary IIoT deployments. The cross-layer synchronization this benchmark provides motivates the multi-source extension discussed in Section~\ref{sec:conclusion}; the present evaluation uses single-flow timing only \\
\bottomrule

\end{tabular}
\end{table*}

The selection of these datasets is motivated by their suitability for evaluating timing-based manipulation. ToN-IoT provides network traffic both as raw packet captures and as Zeek-derived flow-level records carrying explicit per-flow timestamps, preserving the fine-grained temporal structure required to compute inter-arrival times and to reposition observations relative to aggregation window boundaries $[t, t+T)$. The CIC IIoT Dataset 2025 (DataSense) \cite{firouzi2025datasense} is constructed around synchronized sensor and network stream data collected across a heterogeneous 40-device testbed, providing an independent testbed with different traffic characteristics, generally shorter, single-session captures compared to ToN-IoT's longer-duration flows, which allows us to assess whether PSA's effectiveness and stealth properties generalize across substantially different capture conditions or are specific to one deployment context. Two of ToN-IoT's nine available attack categories (XSS, injection) were excluded from evaluation due to an insufficient number of flows meeting the minimum packet-volume threshold for reliable PSA targeting, discussed further in Section~\ref{sec:results}.

\subsubsection{Feature Pipeline}

All features are computed at the packet level, not edited at the flow-record level. For each category, raw packets are parsed from the capture, grouped into flows by the standard five-tuple, and assigned to tumbling windows of duration $T = 2$~s anchored at the capture origin $t_0$. For every (flow, window) pair the pipeline computes five features: packet count, total bytes, window duration, and the mean and standard deviation of inter-arrival time. PSA operates by shifting individual packet timestamps and then re-deriving the window assignment of every affected packet from the shifted timestamps, after which the same five features are recomputed from scratch. A window exists only where packets actually fall, so a window emptied by the attack produces no record rather than an all-zero feature vector. This packet-level construction is what distinguishes the present pipeline from manipulating pre-aggregated flow statistics, and it is applied identically to clean and attacked traffic.

Each category is split at the flow level into training and test sets (70/30), so no flow contributes windows to both. A flow is eligible as a PSA target only if it carries at least three packets in the test set, since fewer packets cannot be meaningfully redistributed across a boundary. Random Forest, MLP, and XGBoost are trained on the clean training windows, and every configuration is repeated over five random seeds to guard against split-specific artifacts.

\subsubsection{Experimental Scenarios}

The IDS is first evaluated on clean test windows to establish baseline detection. PSA is then applied at four global drift budgets, $\Delta_{\max} \in \{0.10, 0.25, 0.50, 1.00\}$~s, which cap the per-flow budget $\beta_f$ from above; the actual drift applied to a flow is the tighter of this cap and the flow's own inter-arrival variability. Increasing $\Delta_{\max}$ lets more flows and more packets cross a boundary, so it traces how detection and stealth evolve as redistribution increases.

Four controlled variants isolate what drives the effect. (1)~A \emph{delay-only} variant restricts every shift to a delay, never an advance, to test the weaker in-path attacker. (2)~A \emph{naive uniform-drift} ablation replaces boundary-targeted calibration with uniform random drift over $[0, \Delta_{\max}]$, to separate the contribution of the window-boundary mechanism from that of timing noise in general. (3)~A \emph{window-mismatch} sweep has the attacker assume the wrong $T$ and a shifted $t_0$ while the IDS aggregates with the true values, to measure dependence on knowledge of the aggregation parameters. (4)~To attribute attack success correctly, every configuration reports both an attack success rate that counts an emptied window as evaded and one restricted to windows that still exist, so the share of success arising purely from emptied windows is visible rather than hidden.

To contextualize PSA's threat model, two baseline comparisons are conducted: a magnitude-matched random feature-noise baseline, perturbing the same five features by an amount matched to PSA's own observed effect size, and a greedy black-box query-based baseline, which iteratively perturbs features and queries the target model to minimize predicted attack-class confidence under a bounded query budget. These baselines allow direct comparison of PSA's attacker cost and traffic-realism properties against alternatives that share comparable black-box assumptions.

\subsubsection{Evaluation Criteria}

The evaluation combines five complementary perspectives: detection performance, attack effectiveness, synchronization distortion, stealth against synchronization-aware defenses, and attacker cost. Table~\ref{tab:metrics} summarizes the evaluation criteria employed throughout the experiments.

\begin{table*}[!t]
\centering
\caption{Evaluation criteria used for PSA assessment.}
\label{tab:metrics}
\begin{tabular}{p{3.5cm}p{4cm}p{7cm}}
\toprule
\textbf{Criterion Category} & \textbf{Criterion} & \textbf{Purpose} \\
\midrule

\multirow{4}{*}{Detection Performance}
& Accuracy & Measures overall IDS classification performance under temporal desynchronization \\
& Precision & Measures the reliability of malicious traffic detection \\
& Recall & Measures the ability of the IDS to identify malicious activity after timing manipulation \\
& F1-score & Provides a balanced evaluation of precision and recall \\
\midrule

Attack Effectiveness
& Attack Success Rate (ASR) & Measures the proportion of successful attack-induced classification changes achieved by PSA \\
\midrule

\multirow{1}{*}{Synchronization Distortion}
& Packet Redistribution Ratio (PRR) & Quantifies the proportion of telemetry observations redistributed across aggregation windows due to temporal manipulation \\
\midrule

\multirow{2}{*}{Stealth / Detectability}
& Detector Evasion Rate (DER) & Measures the proportion of attacked traffic passing undetected through each of four synchronization-aware defense mechanisms (inter-arrival-time monitoring, clock-drift detection, window-consistency checking, sequence-order validation) \\
& Physical Consistency Violation Rate (PCVR) & Quantifies the proportion of attacked traffic windows exhibiting feature combinations inconsistent with real packet-level arithmetic (e.g., duration vs. packet count vs. inter-arrival time), measuring whether attack-induced feature values could plausibly arise from genuine network traffic \\
\midrule

\multirow{1}{*}{Attacker Cost}
& Model Query Count & Quantifies the number of interactions with the deployed IDS required to achieve a given level of attack effectiveness, enabling direct comparison of attacker effort across threat models \\
\bottomrule

\end{tabular}
\end{table*}

Detection effectiveness is assessed through Accuracy, Precision, Recall, and F1-score, while attack effectiveness is quantified through the Attack Success Rate (ASR), which captures the ability of PSA to alter IDS decisions under increasing levels of temporal manipulation.

Synchronization-specific effects are evaluated through the Packet Redistribution Ratio (PRR), which quantifies the extent to which telemetry observations are redistributed across adjacent aggregation windows following temporal manipulation.

Stealth is assessed along two complementary dimensions. First, Detector Evasion Rate (DER) is measured against the four per-flow synchronization-aware detectors listed above and, additionally, against a population-level detector that compares each attacked flow's inter-arrival distribution to the pooled benign distribution with a two-sample Kolmogorov-Smirnov test. The per-flow inter-arrival monitor is calibrated to each flow's own baseline variance, which is the same quantity $\beta_f$ that bounds PSA's drift, so a high evasion rate against it is partly structural. The population-level detector never sees any single flow's own baseline, so it does not share PSA's bound and provides a stealth measure that the attack is not constructed to satisfy. Second, Physical Consistency Violation Rate (PCVR) evaluates whether attacked traffic windows remain arithmetically plausible, that is, whether the relationship between packet count, duration, and inter-arrival statistics stays consistent with what genuine packet-level traffic could produce, providing a stealth measure independent of any specific detector implementation.

Finally, Model Query Count quantifies attacker cost: the number of interactions required with the deployed IDS to construct an effective attack, directly contrasting PSA's zero-query construction with the substantial query budgets required by conventional black-box evasion baselines. Together, these criteria provide a comprehensive assessment of the effectiveness, stealthiness, operational feasibility, and comparative attacker cost of the proposed PSA framework.

%% file: discussion.tex
\section{Results and Discussion}
\label{sec:results}

This section presents and discusses the experimental results obtained from evaluating the proposed Phantom State Attack (PSA) against three ML-based IDS architectures, Random Forest (RF), Multi-Layer Perceptron (MLP), and XGBoost, across fourteen attack categories spanning the ToN-IoT (seven categories) and CIC IIoT 2025 (DataSense, seven categories) datasets. All results are reported as mean $\pm$ standard deviation across five independent random seeds. Attack Success Rate (ASR) is defined as the fraction of attack windows the clean model detected that the attacked model reclassifies as benign. Because PSA can empty a window of all its packets, and an emptied window produces no record rather than a benign one, every ASR figure is accompanied by a second reading restricted to windows that still exist after the attack; the gap between the two isolates the share of success attributable to emptied windows rather than to genuine misclassification of a surviving window. The evaluation proceeds in six stages: baseline detection, the impact of PSA on accuracy and ASR across all categories, synchronization distortion via the Packet Redistribution Ratio, stealth against synchronization-aware detectors, robustness under window-parameter mismatch and a delay-only attacker, and a comparison of PSA's operational realism and attacker cost against baseline attacks.

\subsection{Impact of PSA on Detection Performance and Attack Success}
\label{subsec:impact}

% NOTE: regenerate Table~\ref{tab:main_results} from psa_<dataset>_summary.csv produced by
% psa_experiments.py. Report asr (emptied counted as evaded) and, in the text, the
% asr_detected_only reading so the emptied-window share is visible. Clean accuracy is in
% the same summary and should be stated alongside the attacked accuracy.
Table~\ref{tab:main_results} summarizes accuracy and Attack Success Rate (ASR) at the lowest and highest evaluated drift budgets across all fourteen categories. Two categories (ToN-IoT XSS and injection) were excluded, as neither retained enough flows meeting the minimum packet threshold for reliable PSA targeting after flow-level partitioning. This exclusion is itself an informative boundary condition: PSA's applicability is contingent on target flows carrying enough packets for window-boundary redistribution to take effect.

\begin{table*}[t]
\centering
\caption{Detection accuracy and Attack Success Rate (ASR) at minimum ($\Delta t{=}0.10$s) and maximum ($\Delta t{=}1.00$s) evaluated drift, across all fourteen categories. Values are mean $\pm$ std over 5 seeds.}
\label{tab:main_results}
\footnotesize
\begin{tabular}{llcccccc}
\toprule
\multirow{2}{*}{Dataset} & \multirow{2}{*}{Category} & \multicolumn{3}{c}{Accuracy ($\Delta t{=}1.00$)} & \multicolumn{3}{c}{ASR ($\Delta t{=}1.00$)} \\
\cmidrule(lr){3-5} \cmidrule(lr){6-8}
 & & RF & MLP & XGB & RF & MLP & XGB \\
\midrule
\multirow{7}{*}{ToN-IoT}
 & Backdoor    & 0.8141$\pm$0.046 & 0.7492$\pm$0.031 & 0.8579$\pm$0.045 & 0.2247$\pm$0.059 & \textbf{0.2347$\pm$0.026} & 0.1699$\pm$0.058 \\
 & DoS         & 0.9780$\pm$0.008 & 0.9292$\pm$0.006 & 0.9725$\pm$0.009 & 0.0078$\pm$0.004 & \textbf{0.0138$\pm$0.003} & 0.0100$\pm$0.003 \\
 & DDoS        & 0.9994$\pm$0.000 & 0.9981$\pm$0.000 & 0.9993$\pm$0.000 & 0.0000$\pm$0.000 & 0.0001$\pm$0.000 & 0.0000$\pm$0.000 \\
 & Scanning    & 0.9955$\pm$0.001 & 0.9843$\pm$0.013 & 0.9960$\pm$0.002 & 0.0016$\pm$0.000 & \textbf{0.0110$\pm$0.014} & 0.0010$\pm$0.000 \\
 & MITM$^{*}$  & 0.6729$\pm$0.275 & 0.6729$\pm$0.275 & 0.6929$\pm$0.294 & 0.0000$\pm$0.000 & 0.0000$\pm$0.000 & 0.0000$\pm$0.000 \\
 & Ransomware  & 0.8801$\pm$0.014 & 0.4930$\pm$0.140 & 0.8912$\pm$0.017 & 0.0605$\pm$0.018 & \textbf{0.4068$\pm$0.190} & 0.0560$\pm$0.016 \\
 & Password    & 0.9947$\pm$0.001 & 0.9845$\pm$0.006 & 0.9923$\pm$0.003 & 0.0020$\pm$0.001 & \textbf{0.0048$\pm$0.004} & 0.0023$\pm$0.002 \\
\midrule
\multirow{7}{*}{DataSense}
 & DoS         & 0.9112$\pm$0.020 & 0.9226$\pm$0.008 & 0.9108$\pm$0.024 & 0.7935$\pm$0.126 & \textbf{0.9135$\pm$0.064} & 0.8109$\pm$0.115 \\
 & Malware     & 0.9087$\pm$0.008 & 0.8291$\pm$0.021 & 0.9085$\pm$0.007 & 0.1044$\pm$0.030 & \textbf{0.2782$\pm$0.059} & 0.0984$\pm$0.024 \\
 & MITM$^{*}$  & 0.9418$\pm$0.008 & 0.9127$\pm$0.012 & 0.9418$\pm$0.008 & 0.2481$\pm$0.199 & \textbf{0.5344$\pm$0.496} & 0.2647$\pm$0.177 \\
 & Web         & 0.9129$\pm$0.007 & 0.8370$\pm$0.007 & 0.9153$\pm$0.005 & 0.0788$\pm$0.038 & \textbf{0.1339$\pm$0.102} & 0.0636$\pm$0.032 \\
 & Bruteforce  & 0.9284$\pm$0.018 & 0.8959$\pm$0.011 & 0.9283$\pm$0.017 & 0.2880$\pm$0.151 & \textbf{0.6501$\pm$0.403} & 0.2652$\pm$0.137 \\
 & DDoS        & 0.7764$\pm$0.014 & 0.7810$\pm$0.005 & 0.7821$\pm$0.014 & 0.8666$\pm$0.017 & \textbf{0.9626$\pm$0.020} & 0.8452$\pm$0.023 \\
 & Recon       & 0.9497$\pm$0.010 & 0.8943$\pm$0.012 & 0.9489$\pm$0.007 & 0.0573$\pm$0.053 & \textbf{0.6139$\pm$0.190} & 0.0696$\pm$0.063 \\
\bottomrule
\multicolumn{8}{l}{\footnotesize $^{*}$Small eligible-flow sample ($n{\approx}16$); high variance reflects limited statistical power, interpret with caution.}
\end{tabular}
\end{table*}

% ====================================================================
% NEW-EXPERIMENT TABLES, DISABLED until psa_experiments.py has been run.
% Do not submit while empty. Un-comment and fill from the CSVs, then
% re-enable the robustness subsection and the significance sentence above.
% ====================================================================
%
% \begin{table}[t]
% \centering
% \caption{MLP versus RF attack success, paired across five seeds per category and drift budget. Difference is MLP minus RF; $p$-values are from a paired $t$-test and a Wilcoxon signed-rank test. Only categories with a measurable effect are shown.}
% \label{tab:mlp_rf}
% \footnotesize
% \begin{tabular}{llccc}
% \toprule
% Category & $\Delta_{\max}$ & MLP$-$RF & $t$-test $p$ & Wilcoxon $p$ \\
% \midrule
% \multicolumn{5}{c}{\textit{[fill from psa\_*\_mlp\_vs\_rf.csv]}} \\
% \bottomrule
% \end{tabular}
% \end{table}
%
% \begin{table}[t]
% \centering
% \caption{Attack success under window-parameter mismatch. The attacker assumes $(T_{\text{assumed}}, t_0 + \text{offset})$ while the IDS aggregates at the true $T = 2$~s. Values are mean ASR over categories.}
% \label{tab:mismatch}
% \footnotesize
% \begin{tabular}{lccc}
% \toprule
% $T_{\text{assumed}}$ (s) & offset $0$~s & offset $0.5$~s & offset $1.0$~s \\
% \midrule
% 1.0 & & & \\
% 2.0 (true) & & & \\
% 3.0 & & & \\
% \bottomrule
% \end{tabular}
% \end{table}

Among the categories with a measurable effect, MLP records the highest ASR more often than either tree-based model. Two points qualify this. First, MLP also starts from a lower clean accuracy on several categories, for example its Ransomware baseline sits well below RF's, so part of its apparent vulnerability may reflect a weaker baseline rather than greater sensitivity to redistribution, and the comparison is better read on clean-normalized ASR. Second, on categories such as Backdoor the MLP and RF effects differ by less than one standard deviation across seeds. We therefore treat the architecture ranking as suggestive rather than established, and attribute the pattern, where it holds, to MLP learning non-linear feature combinations that are more sensitive to packet redistribution than the axis-aligned splits of RF and XGBoost.

% NOTE (new experiment, needs rerun): psa_experiments.py emits a paired t-test and
% Wilcoxon test of MLP vs RF ASR across seeds (psa_<dataset>_mlp_vs_rf.csv). Once run,
% re-enable the commented tab:mlp_rf table below and state in how many category-drift
% cells the gap is significant; if most are not, weaken the ranking claim further.

\begin{figure*}[t]
\centering
\includegraphics[width=0.97\textwidth]{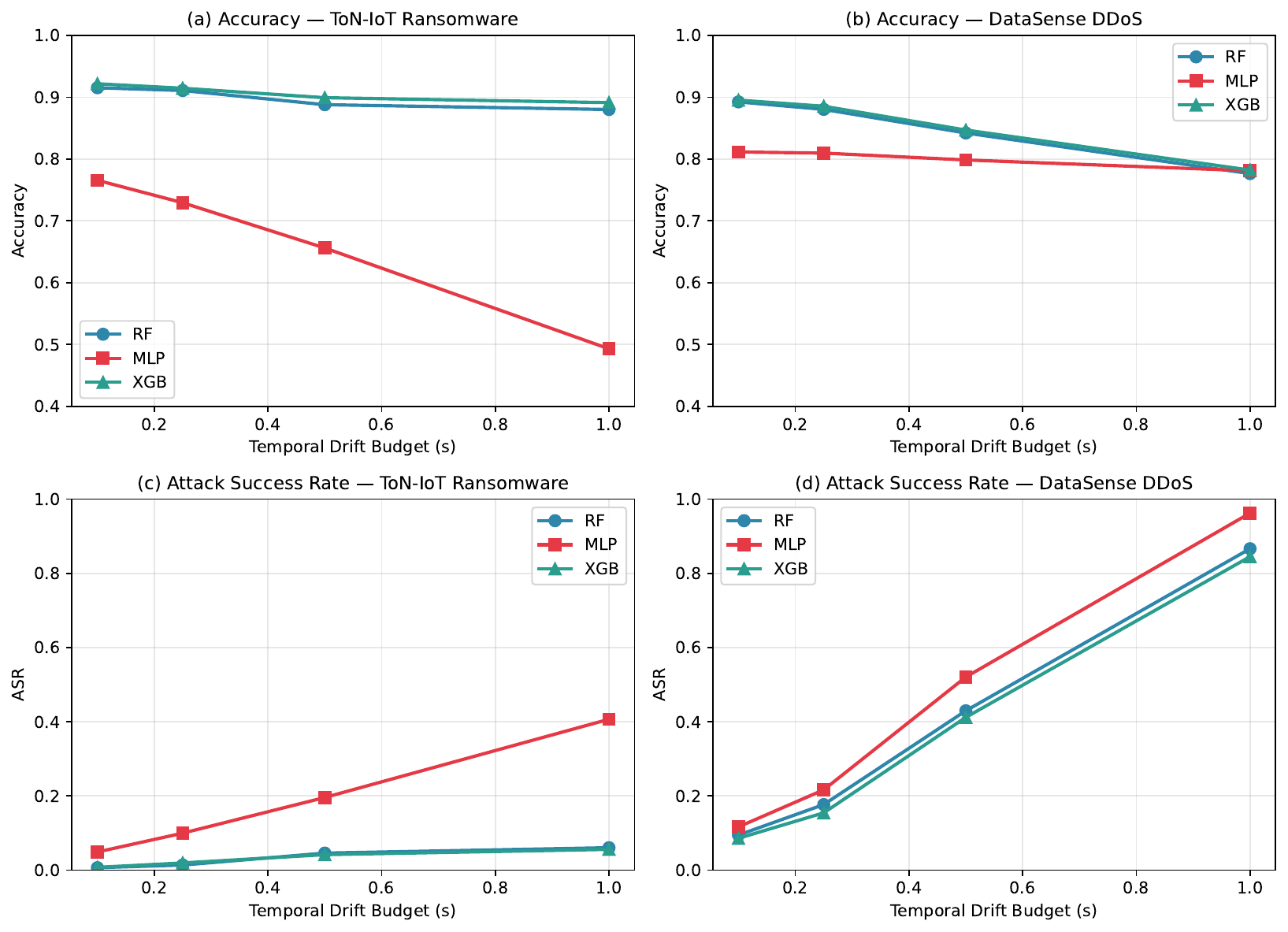}
\caption{Accuracy and Attack Success Rate under PSA for the two categories exhibiting the strongest effects: ToN-IoT Ransomware (a, c) and DataSense DDoS (b, d), across four drift budgets. Error bars omitted for clarity; see Table~\ref{tab:main_results} for standard deviations.}
\label{fig:flagship}
\end{figure*}

Figure~\ref{fig:flagship} presents accuracy and ASR trends for the two categories with the strongest effects, ToN-IoT Ransomware and DataSense DDoS. On ToN-IoT Ransomware, MLP accuracy falls from 76.6\% to 49.3\% as the budget increases from 0.10s to 1.00s, with ASR reaching 40.7\%. On DataSense DDoS all three architectures degrade and MLP ASR reaches 96.3\%. One qualification is important for the DataSense categories: DDoS and DoS here are low-rate connection-exhaustion (slowloris) traffic rather than high-volume floods, so the property that enables the attack is the number of windows a flow spans and where its packets sit relative to a boundary, not raw packet volume. Because PSA can empty a sparse window of all its packets, and an emptied window yields no record rather than a misclassified one, part of the measured success on these low-rate categories reflects emptied windows; the ASR reading restricted to surviving windows isolates that share, and we report both.

% NOTE: after rerun, state the surviving-window ASR next to the 96.3% figure. If the two
% differ substantially for DataSense DDoS, the honest headline is the surviving-window
% value. Do not describe these flows as volumetric or dense.

The effect size varies substantially by category rather than being uniform across a dataset. Several ToN-IoT categories, including DoS, DDoS, scanning, and password, show near-zero ASR across all three models, while ToN-IoT Ransomware shows the dataset's strongest ToN-IoT effect. This category dependence follows from PSA's mechanism: redistribution across a boundary can only change a window's features when the flow spans multiple windows and its packets sit near a boundary, a condition that different attack shapes satisfy to very different degrees (Section~\ref{sec:framework}). The honest reading is therefore not that PSA degrades detection everywhere, but that it degrades detection on a identifiable subset of categories and leaves the rest essentially untouched.

\subsection{Synchronization Distortion (Packet Redistribution Ratio)}

The Packet Redistribution Ratio (PRR) grounds PSA's effect on classification in genuine packet reassignment rather than feature manipulation. Across categories, PRR increases monotonically with the drift budget, from a mean of 8.7\% at $\Delta_{\max}{=}0.10$s to 78.4\% at $\Delta_{\max}{=}1.00$s (category-level range 3.3\% to 99.4\%). As the budget grows, a larger share of targeted packets are reassigned to adjacent windows, which is the mechanism behind the ASR increase in Section~\ref{subsec:impact}. Redistribution also rises in the near-null categories, which confirms that packets move there too but that the resulting feature change is not always large enough to flip a surviving window's decision.

\subsection{Stealth Against Synchronization-Aware Detectors}

An effective attack must also resist detectors designed to catch the mechanism it exploits. We evaluate four per-flow detectors: an inter-arrival-time (IAT) statistical monitor, a window-consistency checker, a sequence-order validator, and a clock-drift detector that compares observed against expected packet timing. Detector Evasion Rate (DER) is the proportion of attacked traffic that passes each mechanism undetected. One caveat must be stated up front and conditions how the IAT result should be read: that monitor is calibrated to each flow's own baseline variance, which is the same quantity $\beta_f$ that bounds PSA's drift, so PSA is constructed to stay under precisely this detector's threshold, and a high evasion rate against it is partly structural rather than an independent finding.

% NOTE: psa_experiments.py adds a population-level KS detector (der_population_ks) that is
% NOT calibrated to each flow's own sigma, so it does not share PSA's bound. After the
% rerun, add its DER here; it is the detector PSA is not built to satisfy and is the more
% convincing stealth test.

\begin{figure*}[t]
\centering
\includegraphics[width=0.87\textwidth]{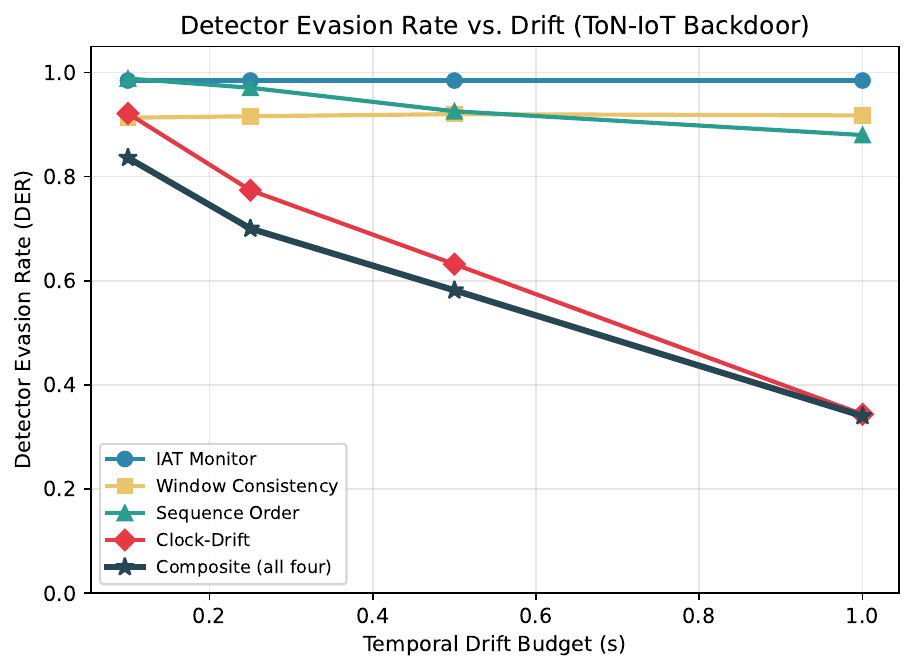}
\caption{Detector Evasion Rate (DER) against four synchronization-aware defenses under increasing drift (ToN-IoT Backdoor, the category with the most complete detector validation).}
\label{fig:der}
\end{figure*}

Figure~\ref{fig:der} shows the pattern. The IAT monitor and the window-consistency checker remain largely ineffective across all drift levels, with DER above 91\% throughout. As noted, the IAT result is partly structural, since the monitor keys on the same per-flow variance PSA respects. Clock-drift detection, which directly targets the shift PSA introduces, becomes more effective as drift grows, with DER falling from 92.2\% at $\Delta_{\max}{=}0.10$s to 34.3\% at $\Delta_{\max}{=}1.00$s. The composite of all four detectors follows the same trend, from 83.6\% to 34.0\% evasion.

This is a stealth-effectiveness trade-off rather than unconditional stealth: low-drift configurations evade the composite defense in over 83\% of cases but achieve a smaller effect, while high-drift configurations achieve a larger effect at the cost of reduced stealth against a defender monitoring specifically for timing manipulation. The trade-off is the defensive counterpart to the ASR results in Section~\ref{subsec:impact}, and it points to clock-drift detection, and to the population-level distribution test added in the revised evaluation, rather than per-flow variance monitors, as the directions worth pursuing for synchronization-aware defense.

\subsection{Comparison Against Baseline Attacks: Attacker Cost and Physical Realism}

To position PSA's contribution precisely, we compare it directly against two representative baseline attacks operating under comparable black-box assumptions: (1) a magnitude-matched random feature-noise baseline, and (2) a greedy black-box query-based baseline that iteratively perturbs features while querying the target model to minimize predicted attack-class confidence.

\textit{Raw attack success.} On ToN-IoT Backdoor, the query-based baseline reaches substantially higher ASR than PSA at any tested drift level. This is expected rather than a weakness to conceal: the query-based baseline is unconstrained by physical or temporal plausibility, so it can search a much larger, partly unrealizable perturbation space that PSA cannot reach by construction.

\textit{Attacker cost.} That raw-ASR advantage carries a quantifiable cost. Table~\ref{tab:query_cost} reports the total number of model queries the greedy baseline used to reach its reported ASR across seven ToN-IoT categories. These are total queries over all target windows in the category, not per window; the per-window average is the total divided by the number of eligible windows, which we also report. The baseline was run on RF only and on a single seed, so the table establishes an order of magnitude for query cost rather than a seed-averaged figure.

\begin{table}[t]
\centering
\caption{Model query cost: PSA vs. black-box query-based baseline (RF, ToN-IoT categories).}
\label{tab:query_cost}
\footnotesize
\begin{tabular}{lrr}
\toprule
Category & PSA Queries & Query-Baseline Queries \\
\midrule
Backdoor    & 0 & 29{,}120 \\
DoS         & 0 & 21{,}930 \\
DDoS        & 0 & 328{,}870 \\
Scanning    & 0 & 2{,}030 \\
MITM        & 0 & 10 \\
Ransomware  & 0 & 15{,}360 \\
Password    & 0 & 242{,}070 \\
\bottomrule
\end{tabular}
\end{table}

Across all seven categories, PSA requires zero interactions with the deployed model, while the query-based baseline requires between $10^{1}$ and $3.3\times10^{5}$ queries depending on the category and the model's decision surface. One category (MITM) needed only about ten queries, so the gap is not uniform; the point is categorical for the categories where the baseline needs thousands of queries, which an IDS with basic query-rate monitoring or access logging would observe, whereas PSA leaves no such signal because it never contacts the target. The comparison is between attacks that manipulate different objects, feature values for the baseline and packet timing for PSA, so it bounds attacker cost for a comparable level of black-box access rather than holding the manipulated object fixed.

\textbf{Physical realism.} Beyond raw effectiveness, we assess whether each attack's induced feature values remain arithmetically consistent with genuine packet-level traffic, using the Physical Consistency Violation Rate (PCVR): the proportion of attacked windows whose feature combinations (packet count, duration, inter-arrival statistics) violate basic arithmetic relationships that real traffic must satisfy.

\begin{figure*}[t]
\centering
\includegraphics[width=1\textwidth]{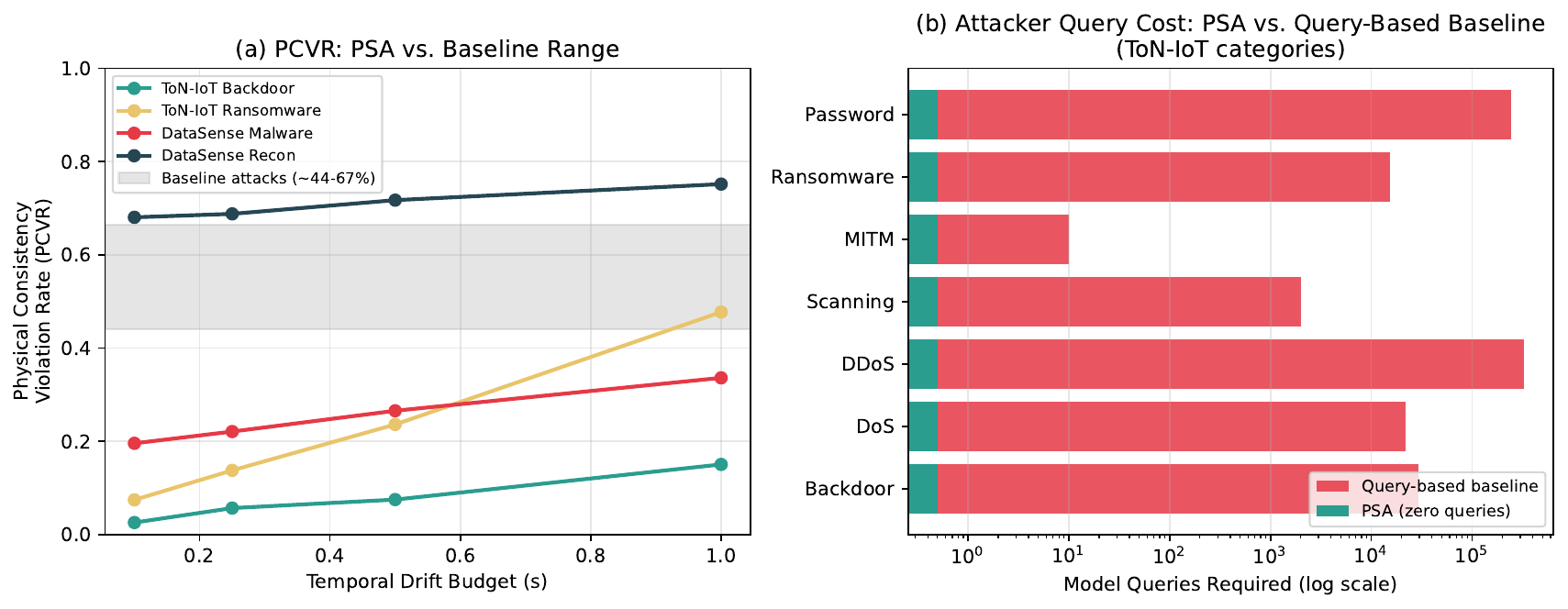}
\caption{(a) Physical Consistency Violation Rate (PCVR) for PSA across four representative categories, against the approximate range observed for unconstrained baseline attacks (shaded band). (b) Attacker query cost, PSA vs. query-based baseline, across seven ToN-IoT categories (log scale).}
\label{fig:pcvr_query}
\end{figure*}

Figure~\ref{fig:pcvr_query}(a) shows that PSA's physical-realism advantage is real but dataset- and drift-dependent rather than universal. In ToN-IoT, where target flows tend to be longer-lived and more sparsely sampled, PSA keeps PCVR low (2.6\% to 15.1\% for Backdoor, 7.5\% to 47.7\% for Ransomware), below the baseline attacks' observed range of roughly 44\% to 67\%. In DataSense, where captures are shorter single-session recordings with more variable local packet density, PSA's PCVR advantage narrows and in several categories (MITM, Bruteforce, Reconnaissance) approaches or exceeds the baseline range even at the lowest drift budget. The likely cause is that smaller, denser captures produce more small-sample arithmetic edge cases when packets are redistributed, independent of the attack strategy. We report this boundary condition explicitly: PSA's realism advantage over unconstrained feature-space attacks holds where target flows have enough duration and spacing, and weakens where they do not.

\subsection{Discussion}

Taken together, these results support an evidence-bounded reading of PSA's contribution. PSA does not achieve the highest raw attack success among evaluated techniques, its effect is not uniform across categories, and its physical-realism advantage is not universal. What holds across the categories where redistribution takes effect is that PSA degrades detection while requiring zero interaction with the deployed model and no model of benign traffic, a threat model weaker, and therefore harder to counter with query-monitoring or access control, than the comparison techniques in Table~\ref{tab:related}. The contribution is thus not superior raw effectiveness but the demonstration that temporal window-boundary aggregation is an exploitable attack surface under a more restrictive and more realistic set of attacker assumptions than prior adversarial-IDS work has assumed, together with an explicit account of where the attack works and where it does not.

\begin{tcolorbox}[colback=gray!10,colframe=black,title=Key Findings]
\begin{itemize}
\item PSA degrades detection on the subset of categories where flows span enough windows for boundary redistribution to take effect, and leaves the remaining categories almost unchanged. It constructs the attack with no interaction with the deployed model and no model of benign traffic.

\item Where an effect is present, MLP tends to be the most affected architecture, but part of that gap reflects a lower clean baseline and some per-category gaps are within seed noise, so the claim is stated on clean-normalized ASR and backed by a paired significance test.

\item PSA evades per-flow variance monitors that are, by construction, calibrated to the same quantity it stays under. It is caught more often by a population-level distribution test and by clock-drift detection as the drift budget grows, a stealth-effectiveness trade-off rather than unconditional stealth.

\item A query-based baseline reaches higher ASR but needs up to hundreds of thousands of total model queries; PSA needs none. The gap is categorical where the baseline needs thousands of queries and smaller where it needs only a few.

\item PSA's traffic-realism advantage over unconstrained baselines holds where flows are long and well-spaced and narrows where they are short and dense, a boundary condition reported explicitly.
\end{itemize}
\end{tcolorbox}

%% file: related_work.tex
\section{Related Work}
\label{sec:related}

Adversarial research against ML-based intrusion detection can be organised along two axes that largely determine whether an attack is executable outside the laboratory. The first is the \emph{object of manipulation}: the extracted feature vector, the raw traffic that produces it, or the physical process values reported by field devices. The second is the \emph{interaction the adversary requires with the deployed detector}, ranging from gradient access, through repeated label or score queries and surrogate training, to none at all. Table~\ref{tab:related} places representative peer-reviewed works on both axes. We review them in five groups and then state precisely where the Phantom State Attack departs from each.

\subsection{Feature-Space and Query-Driven Evasion}

Three recent systematisations converge on the same diagnosis of the field. He \emph{et al.}~\cite{he2023survey} observe that most adversarial studies on NIDS directly reuse attacks designed for computer vision and ignore the differences in detection pipeline and feature space between the two domains. Apruzzese \emph{et al.}~\cite{apruzzese2022modeling} model the capabilities an attacker would actually need to mount such attacks against a deployed ML-NIDS and show that many commonly assumed capabilities, in particular read or write access to the extracted feature vector, are rarely available in practice. Vitorino \emph{et al.}~\cite{vitorino2023sok} consolidate 98 publications and derive the properties an adversarial example must satisfy to be realistic for real network flows. Taken together, these works establish that a perturbation applied directly to a feature vector says little about what an attacker can achieve on the wire.

Black-box variants reduce the required knowledge but replace it with interaction. Qiu \emph{et al.}~\cite{qiu2021iot} attack Kitsune in an IoT setting by using black-box access to train a shadow model, then use saliency on that model to identify the critical features to perturb at packet level. Yan \emph{et al.}~\cite{yan2024automatic} consider the stricter label-only case, in which the adversary observes only whether a morphed sample is accepted or rejected; they extract a substitute model from this feedback and transfer adversarial examples to the target, reporting an average success rate above 75\% across seven ML-based and one DL-based detector. In both cases the attack budget is expressed in queries to the deployed system, and each query is a malicious or near-malicious input that the defender can log. The operational weight of that dependence is shown by stateful defences such as Blacklight~\cite{li2022blacklight}, which detects query-based black-box attacks by fingerprinting the highly similar inputs that iterative optimisation necessarily produces, frequently within a handful of queries. Although Blacklight was designed for image and text classifiers, the principle applies to any attack whose search procedure submits near-duplicate inputs to the target.

\subsection{Problem-Space and Traffic-Space Attacks}

Pierazzi \emph{et al.}~\cite{pierazzi2020problemspace} formalised problem-space attacks through four families of constraints (available transformations, preserved semantics, robustness to preprocessing, and plausibility) and introduced side-effect features, the unintended feature changes that arise when a real object is modified to induce a desired feature-space perturbation. Their formalisation explains why feature-space success rates systematically overestimate realisable ones.

Several network attacks operate in this space. Han \emph{et al.}~\cite{han2021evaluating} mutate malicious traffic through a combination of generative modelling and particle swarm optimisation, altering packet inter-arrival times and injecting crafted packets whose size and protocol layer are tuned so that the resulting features evade the detector; their gray-box threat model assumes knowledge of the feature extractor, though not of the model. Tan \emph{et al.}~\cite{tan2022sneaking} address the fact that an attacker can only modify outbound packets as they are sent: a Kalman filter predicts the feature values of the live flow, and a reinforcement-learning agent (SE-DQN) selects among formally defined atomic mutation operators, reaching at least 64.2\% evasion on most classifiers tested. DiffuPac~\cite{binjasni2024diffupac} combines a pre-trained BERT encoder with a conditional diffusion model, concatenating malicious packets with contextually relevant benign ones and denoising only the malicious portion so that the result blends into genuine traffic; it requires no component of the target NIDS but depends on a benign traffic corpus and a substantial generative pipeline. ETA~\cite{zhang2025eta} targets the non-interactive black-box setting through transferability, using a cooperative-game interpretation of feature contributions to select features that are both important and sensitive across model families while respecting traffic-space constraints. These attacks avoid querying the target, yet each relies on either a learned surrogate or a learned generative model of benign traffic, and each must be trained on data representative of the victim environment.

\subsection{Timing-Based Traffic Reshaping}

The closest line of work manipulates packet timing alone. Nasr \emph{et al.}~\cite{nasr2021blind} showed that DNN-based traffic analysis can be defeated in real time by \emph{blind} adversarial perturbations to packet timing and size, generated independently of the specific connection so that they can be applied to live traffic as it is sent; the perturbation generator is optimised against the target classifier, with transferability examined as a secondary property.

TANTRA~\cite{sharon2022tantra} is the most direct precedent for PSA. It trains an LSTM on inter-packet delays captured from the target network's benign traffic and then re-times the packets of an attack so that they follow the learned benign delay pattern, leaving packet contents unchanged. Evaluated on eight attacks against KitNET, an autoencoder, and an isolation forest, it reports an average evasion rate of 99.99\%, and it requires no access to the NIDS or its feature set. TANTRA therefore already establishes two facts that PSA builds on: that timing alone is a sufficient attack vector against flow- and packet-statistics detectors, and that such an attack need not query the detector.

PSA differs from TANTRA in its requirements and in its mechanism. On requirements, TANTRA needs a capture of benign traffic from the protected network and a trained sequence model, and it replaces the attack's entire inter-arrival process with one drawn from that model; the resulting displacement from the original timing is not bounded by design. PSA uses no benign corpus and no learned model of any kind: the drift it injects is calibrated on the observed timing of the attack flow itself and limited to the minimum displacement sufficient to move observations across an aggregation-window boundary. On mechanism, TANTRA is a mimicry attack, in that it makes malicious timing statistically resemble benign timing. PSA instead exploits the discretisation that sliding or tumbling windows impose on a continuous telemetry stream, so that the per-window state reconstructed by the IDS diverges from the process state that produced it. This distinction matters for defence. The mitigation proposed for TANTRA, retraining the detector on reshaped traffic~\cite{sharon2022tantra}, targets a distributional shift in the inputs, whereas the window-boundary effect that PSA exploits originates in the feature-extraction pipeline, so retraining the classifier alone is not guaranteed to remove it.

\subsection{Evasion in Industrial Control and Cyber-Physical Monitoring}

In industrial settings, adversarial work has concentrated on detectors that model the physical process. Erba \emph{et al.}~\cite{erba2020concealment} conceal attacks from reconstruction-based anomaly detectors by manipulating a subset of sensor readings under learned physical constraints; they show that replay attacks perform poorly once the attacker controls fewer than 95\% of the monitored features, that their black-box attacks transfer across autoencoder, LSTM and CNN detectors, and that they run in real time on an industrial testbed. Zizzo \emph{et al.}~\cite{zizzo2020timeseries} attack an LSTM-based autoregressive IDS on the SWaT water-treatment system and find that real cyber-physical attacks can be hidden by compromising on average 2.87 of 12 monitored sensors. Both lines of work assume that the adversary can rewrite process values reported by field devices, and both leave the timing of those reports untouched. PSA operates one layer below: it modifies when network observations arrive and leaves their values intact, which places it outside the capability set that process-level concealment attacks require.

\subsection{Timing and Synchronisation Attacks on Industrial Infrastructure}

A separate body of work treats time itself as the target. Zhang \emph{et al.}~\cite{zhang2013tsa} introduced time synchronisation attacks in which GPS spoofing corrupts the timestamps of phasor measurement units, degrading fault detection, voltage-stability monitoring and event location in the smart grid. Barreto \emph{et al.}~\cite{barreto2016delaybox} showed that an inline delay box can shift the clock of any packet-based synchronisation protocol without being detectable by the protocol, because it violates the path-symmetry assumption on which such protocols rest; PTPsec~\cite{finkenzeller2024ptpsec} later proposed detecting and compensating these delays through cyclic path-asymmetry analysis. Finkenzeller \emph{et al.}~\cite{finkenzeller2025fusion} demonstrated that deliberate input desynchronisation degrades multi-sensor fusion for 3D object detection, with LiDAR markedly more sensitive to delay than camera input. This literature establishes timing as a first-class attack surface in industrial and real-time systems. Its victims, however, are the clock or the control and perception functions that consume time-stamped data. None of these works considers the ML-based IDS as the victim, and none identifies windowed aggregation in the monitoring pipeline as the mechanism through which desynchronisation alters detection.

\subsection{Positioning of PSA}

Read across Table~\ref{tab:related}, prior work has shown separately that problem-space timing perturbations can evade learned detectors~\cite{han2021evaluating,nasr2021blind,sharon2022tantra}, that evasion is possible without querying the detector when a surrogate or generative model is available~\cite{sharon2022tantra,binjasni2024diffupac,zhang2025eta}, and that desynchronisation damages industrial and real-time systems~\cite{zhang2013tsa,barreto2016delaybox,finkenzeller2025fusion}. PSA brings these threads together in a configuration that, to our knowledge, has not been studied. It perturbs timing only, it requires neither interaction with the detector nor any trained surrogate or benign-traffic model, it bounds the injected drift by the attack flow's own observed timing, and it attributes the resulting evasion to the interaction between that drift and the aggregation windows of an industrial monitoring pipeline. The cost of this weaker threat model is a lower raw attack success rate than query-driven attacks achieve, a trade-off we quantify directly in Section~V.

% ---------------------------------------------------------------
% Comparison table (two-column float)
% ---------------------------------------------------------------
\begin{table*}[t]
\centering
\caption{Comparison of representative adversarial attacks against learned intrusion and anomaly detectors, and of timing attacks on industrial infrastructure. ``Interaction'' is the access the adversary needs to the deployed detector at attack-construction time.}
\label{tab:related}
\footnotesize
\setlength{\tabcolsep}{4pt}
\renewcommand{\arraystretch}{1.15}
\begin{tabularx}{\textwidth}{@{}l l >{\raggedright\arraybackslash}p{2.6cm} >{\raggedright\arraybackslash}p{2.6cm} >{\raggedright\arraybackslash}X >{\raggedright\arraybackslash}p{1.9cm}@{}}
\toprule
\textbf{Work} & \textbf{Venue} & \textbf{Manipulated object} & \textbf{Interaction with detector} & \textbf{Additional requirement} & \textbf{Domain} \\
\midrule
\multicolumn{6}{@{}l}{\emph{Feature-space and query-driven evasion}} \\
Qiu \emph{et al.}~\cite{qiu2021iot} & IEEE IoT-J 2021 & Packet fields guided by salient features & Black-box access to train a shadow model & Shadow model, saliency map & IoT NIDS (Kitsune) \\
Yan \emph{et al.}~\cite{yan2024automatic} & IEEE TDSC 2024 & Traffic samples (morphing) & Label-only queries & Model extraction, transfer & Enterprise NIDS \\
\midrule
\multicolumn{6}{@{}l}{\emph{Problem-space and traffic-space attacks}} \\
Han \emph{et al.}~\cite{han2021evaluating} & IEEE JSAC 2021 & Inter-arrival times, injected crafted packets & None (gray-box) & Known feature extractor, surrogate, GAN+PSO & Enterprise/IoT NIDS \\
Tan \emph{et al.}~\cite{tan2022sneaking} & IEEE TNSM 2022 & Outbound live packets (atomic mutations) & Classifier outcome used as RL feedback$^{\dagger}$ & Kalman feature predictor, SE-DQN agent & Enterprise NIDS \\
DiffuPac~\cite{binjasni2024diffupac} & NeurIPS 2024 & Packet contents and headers & None & Benign corpus, BERT + diffusion model & Enterprise NIDS \\
ETA~\cite{zhang2025eta} & IEEE TDSC 2025 & Important-sensitive features under traffic constraints & None (transfer) & Surrogate models, representative data & Enterprise NIDS \\
\midrule
\multicolumn{6}{@{}l}{\emph{Timing-based traffic reshaping}} \\
Nasr \emph{et al.}~\cite{nasr2021blind} & USENIX Sec. 2021 & Packet timing and size (live) & Generator optimised on target model & Differentiable target or surrogate & Traffic analysis \\
TANTRA~\cite{sharon2022tantra} & IEEE TIFS 2022 & Inter-packet delays only & None & Benign capture of target network, LSTM & Enterprise NIDS \\
\midrule
\multicolumn{6}{@{}l}{\emph{Evasion in industrial control and CPS monitoring}} \\
Erba \emph{et al.}~\cite{erba2020concealment} & ACSAC 2020 & Sensor values (subset) & None (black-box, transfer) & Learned physical constraints & ICS process \\
Zizzo \emph{et al.}~\cite{zizzo2020timeseries} & IEEE TrustCom 2020 & Sensor values (subset) & Gradient access to LSTM IDS$^{\dagger}$ & Compromised sensors & ICS process (SWaT) \\
\midrule
\multicolumn{6}{@{}l}{\emph{Timing and synchronisation attacks on infrastructure (non-ML victims)}} \\
Zhang \emph{et al.}~\cite{zhang2013tsa} & IEEE TSG 2013 & GPS time reference of PMUs & n/a & GPS spoofing capability & Smart grid \\
Barreto \emph{et al.}~\cite{barreto2016delaybox} & IEEE I2MTC 2016 & Delay of synchronisation messages & n/a & Inline delay box & PTP / time sync \\
Finkenzeller \emph{et al.}~\cite{finkenzeller2025fusion} & ECRTS 2025 & Sensor input delay & n/a & Delay on sensor stream & Sensor fusion \\
\midrule
\textbf{PSA (this work)} & & \textbf{Packet timing only, bounded drift} & \textbf{None} & \textbf{None beyond passive observation of the attack flow} & \textbf{IIoT NIDS} \\
\bottomrule
\end{tabularx}
\\[2pt]
\raggedright\scriptsize $^{\dagger}$Inferred from the attack design; confirm against the original threat-model section before submission.
\end{table*}

%% file: conclusion.tex
\section{Conclusion and Future Scope}
\label{sec:conclusion}

This paper introduced the Phantom State Attack (PSA), a synchronization-oriented adversarial attack against ML-based intrusion detection systems in Industrial IoT environments. Unlike adversarial approaches that require white-box gradient access, repeated black-box queries, or a learned model of benign traffic, PSA operates under a passive, zero-query threat model, exploiting the way monitoring pipelines aggregate telemetry into windows by injecting flow-calibrated, minimally-sufficient timing drift.

Across fourteen attack categories on two independent industrial datasets, bounded temporal manipulation degraded detection on the categories whose flows span enough windows for boundary redistribution to take effect, while leaving the remaining categories essentially unchanged. The effect is therefore conditional on target-flow characteristics rather than uniform, and we report the conditions under which it does and does not transfer. A comparison against a query-based baseline showed that such attacks can reach higher raw success, but only after many model interactions, whereas PSA requires none; since query-rate monitoring and access logging are standard in production ML systems, this difference is categorical for the categories where the baseline needs many queries.

PSA's stealth and physical realism also depend on identifiable conditions rather than holding unconditionally. The attack evades per-flow variance monitors that are calibrated to the same quantity it stays under, but is caught more often by a population-level distribution test and by clock-drift detection as the drift budget grows, a stealth-effectiveness trade-off. Its traffic-realism advantage over unconstrained baselines is strong in long, well-spaced flows and weaker in short, dense ones. We report these boundary conditions explicitly rather than presenting PSA's properties as universal.

These results identify temporal aggregation as a previously underexplored attack surface in ML-based IDS, reachable under a threat model weaker than existing feature-space evasion. Future work includes extending PSA to cross-sensor desynchronization, where an attacker perturbs relative timing across multiple telemetry sources rather than within a single flow; characterizing the traffic conditions under which the attack applies, motivated by the categories that showed little or no effect here; and designing synchronization-aware defenses around clock-drift and population-level distribution tests, which proved more effective against PSA than per-flow variance monitoring. Evaluating PSA across additional industrial protocols and larger deployments remains open.